# Dynamic cavity effect in Topological Insulator $Bi_2Te_3$ based Passive Q-switched Solid State Laser

Yuan-Yao Lin[1], Wei-Heng Song[1], Peng Lee[1], Yi-RanWang[2], Han Zhang[3] and Chao-Kuei Lee[1,*]
1. Department of Photonics, National Sun Yat-sen University, Kaohsiung city, Taiwan, R.O.C.
2. Institute of Crystal Materials Shandong University, Jinan City, China
3. Shenzhen Engineering Laboratory of Phosphorene and Optoelectronics, Key Laboratory of Optoelectronic Devices and Systems of Ministry of Education and Guangdong Province, College of Optoelectronic Engineering, Shenzhen University, Shenzhen 518060, People's Republic of China
Email: chuckcklee@yahoo.com

**Abstract -** Dynamic cavity effect of a Topological Insulator $Bi_2Te_3$ based Passive Q-switched (QS) solid state laser was investigated. With a saturation intensity as low as a few kW, $Bi_2Te_3$ processed by hydrothermal exfoliation serves a sensitive saturable absorber to initiate Qswitching. Without material damage, a transition from QS state to CW state was observed when the pumping strength increases which was explained by the modulation instability spectrum of laser rate equation.

I. INTRODUCTION

Passive Q-switching (PQS) and passive mode-locking (ML) of lasers are the essential techniques for generating giant laser pulses with pulse width ranging from microseconds to femtoseconds. The formation either PQS pulses or passive ML pulses results from dynamic cavity effect in laser systems which have been studied since laser had been invented. Subtle transient behavior has been delineated by theoretical consideration of a mathematical model of rate equations involving atomic population inversion density of gain media, intra-cavity photon density and the population dynamics of absorber if existed [1]. PQS is one of the mature techniques used to produce repetitive burst of coherent photons since its first demonstration by Wood and Schwarz [2] in $CO_2$ laser system. Such a pulsed operation relies on the significant saturation of the absorber inside a laser cavity. Although direct numerical simulation using the mathematical model of coupled rate equations reveals the observed repeated pulsation dynamics, Powell et al. [3] proposed a perturbed approach based on linear stability analysis and reached the same conclusion in the repetition rate as the experimental works done by Karlov *et al.* [4]. In such a pulsed laser, the saturable absorber (SA) is the key component due to its role of starter and stabilization of the dynamic cavity effects [5][6]. Recently, some new nanosheet materials grouped, topological insulators (TIs) featuring their unique phases of spin-orbital coupling associated with the geometry dimensions, have appeared to be potential as SAs due to its nature of surface state like graphene [7][8]. In addition to the surface states of very high saturation intensity, the bulk states in these TI materials offer ultra-sensitive saturation absorptions. After Wen's group first demonstrated topological insulator's great nonlinearity and feasibility for saturated absorption application as mode-locking the fiber laser in 2012 [9][10], TIs, such as $Bi_2Se_3$, $Bi_2Te_3$ and $Sb_2Te_3$, therefore have drawn a lot of attentions in applications for saturation absorbers of pulsed fiber laser[9][11][12][13][14]. In contrast, when compared to fiber laser, solid-state lasers (SSLs) can provide higher power and pulse energy due to their natures of larger mode areas, high thermal conductivity and low undesirable nonlinear effects [15]. In 2013, Zhang et al. reported the first QPS solid-state laser (SSLs) using $Bi_2Se_3$ as saturation absorber [16] that arouses researches in PQS pulsed SSLs based on TIs with various wavelength and various TIs materials in recent years [16] [17][18] [19][20][21][22][23][24] among which Q-switched mode-locking operations have also been reported REF [21]. Regarding to TI as a SA , for example $Bi_2Te_3$, its low saturable intensity favors low-threshold operation for Q-switch laser and even Q-switched ML operation if carefully designed. But ultra-low threshold reduces the tuning range for stable Q-switching operation as a result of unclear and complex dynamical cavity effects at low saturable intensity and degrades the scalability for very high energy pulse and the tolerance in the parameter space to achieve QS, QS-ML and even continuous wave (CW) mode locking operations. Moreover the state of over-saturation is not clear and to the best of our knowledge, has never been investigated. In this work, an abrupt transition from QS state to CW state in a topological insulator $Bi_2Te_3$ based PQS SSL lasing a wavelength of 1064nm was experimentally observed. It was be theoretically explained by the modulation instability of nonlinear laser system.

II. EXPERIMENTS AND RESULT

A QPS SSL laser system was built based on the V-folded configuration illustrated in Fig. 1. The diode-laser pump source is rated at a maximum power of 44 W into a bandwidth of 3 nm (FWHM) around a central wavelength of 808 nm at 25°C. A fiber bundle with a diameter of 0.4 mm and a numerical aperture of 0.2 and coupling optics were used to focus the pump beam into the laser crystal forming a pump beam of 320 μm in diameter on the laser crystal facet. The 3x3x7 mm $Nd^{3+}$:YAG crystal, doped at 0.5 %, was coated for high-reflection at 1064nm (HR; R＞99.8 %) and anti-reflection at 808 nm on pump end faces perpendicular to the laser beam axis. The laser crystal was wrapped with an indium foil and then mounted in a water-cooled copper block. In contrast to linear resonator configuration widely used in solid state passive Q-switched (QS) lasers, V-folded resonator was used deliberately to prevent TIs absorbing the residual pump beam after Nd:YAG crystal, which not only lowers the modulation depth but also cause instability in laser pulse build-up[14]. The pump mirror, $M_1$ was a planar with antireflection (AR) coated at 808 nm and high-reflection (HR) coated at 1064 nm. The folding mirror $M_2$, having a radius of curvature of 200mm, is high-reflective coated at 1064 nm and anti-reflective coated at 808 nm. Regarding to $M_3$, output coupler (OC), is a planar mirror with 10% partial transmission at 1064 nm. The saturable absorber was TI material, hexagonal $Bi_2Te_3$ single crystals with uniform morphology which was mechanically exfoliated first and then processed by hydrothermal exfoliation [25]. The exfoliated $Bi_2Te_3$ sample was deposited on the 0.25 mm thick quartz substrates with 96% transmittance as a SA. Figure 2(a) plots the saturable absorption of the prepared SA which is measured by using laser source of 1-μm wavelength, 1ns pulse duration and 40 kHz repetition rate. The modulation depth of 14.7 %, saturation intensity of 4.6 $kW/cm^2$ and the non-saturable losses of 12.25 % can be obtained by fitting the experiments to a standard model for saturable absorption. This TI absorber of was then inserted between $M_2$ and $M_3$ for avoiding absorption of residual 808 nm pump beam. The stable passively Q-switching operation was achieved with the resonator when the distance between $M_1$ and $M_2$ is 35 mm and the distance between $M_2$ and $M_3$ is 38 mm, respectively.

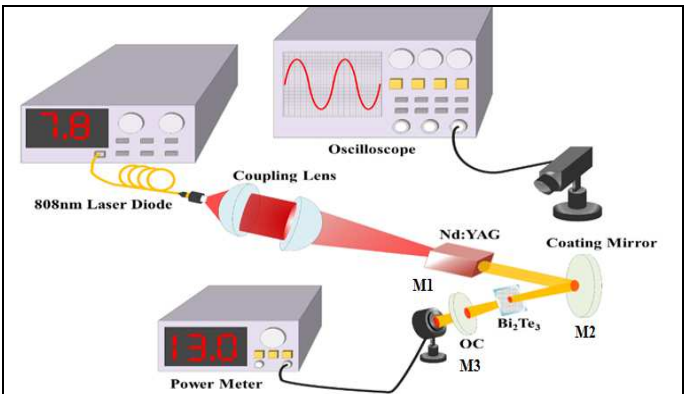

Fig. 1. (color online) Schematic experimental setup of the LD-pumped Nd:YAG laser at 1064 nm.

The near infrared pulse was lased from the SSL when the pump power went above 719.5mW. By continuously increasing the pump power from the threshold and finely tilting the glass substrate, both the pulse repetition rate and the average output power sharply increased, viewing as the typical feature of a passively Q-switched laser. At the pump power of 2W, Fig. 2(b) shows the typical passively Q-switched single pulse profile having a pulse duration of 826 ns and the repetition frequency of the pulse trains is 80.17kHz as illustrated in the inset of Fig. 2(b). Although the average output power grows almost linearly as the pump power increases at a slope rate of 10% as shown in Fig.2(c), the temporal profile of the output laser has a dramatic change at the pump power of 3.5W. For the pump power is under 3.5W, stable QS pulse can be observed in the time trace captured by oscilloscope as illustrated in the inset marked QS state in Fig. 2(c). When the pump power goes above 3.5W pulsed operation becomes unstable. Above the pump power of 4W, the pulsed operation ceases to exist and the measured temporal waveform reveals a stable continuous wave (CW) operation plotted in the inset marked CW state in Fig. 2(c). The spectral intensities of the laser output in both operation regions are measured and are centered at 1064 nm with a very narrow linewidth. Reducing the pump power from above 4W to below 3.5W, the pulsed operation is restored with the output power, repetition frequency and pulse duration being the same as those had been measured previously. By performing the pumping cycle for several times, it is shown that the QS to CW transitions is reversible and repeatable, indicating material damage issue in saburable absorber is void.

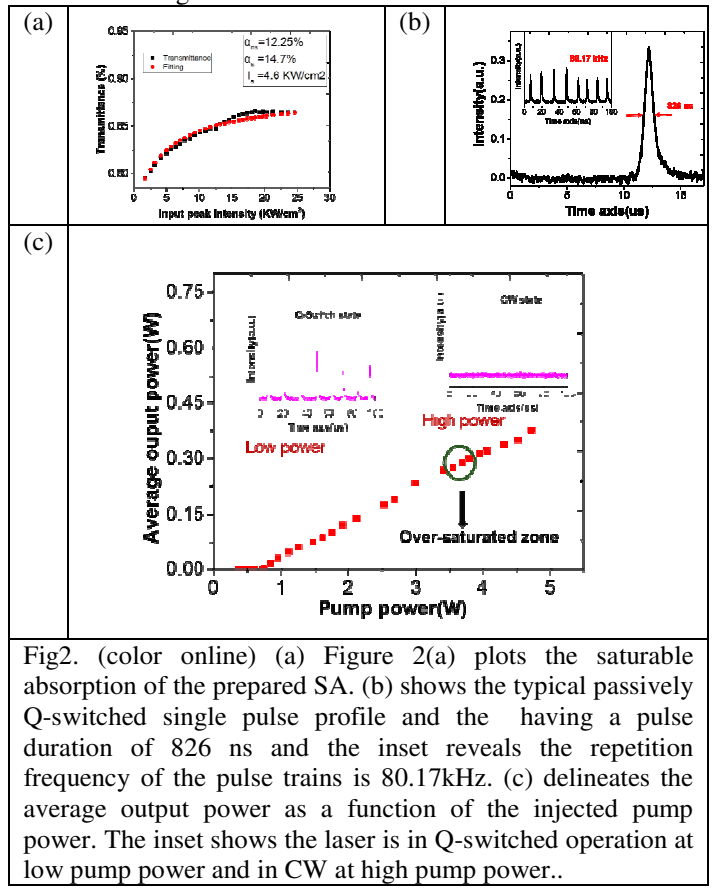

Fig2. (color online) (a) Figure 2(a) plots the saturable absorption of the prepared SA. (b) shows the typical passively Q-switched single pulse profile and the having a pulse duration of 826 ns and the inset reveals the repetition frequency of the pulse trains is 80.17kHz. (c) delineates the average output power as a function of the injected pump power. The inset shows the laser is in Q-switched operation at low pump power and in CW at high pump power..

The observed phenomena should be the effect of over-saturtion, which have been rarely mentioned and have never been discussed before. As mentioned in literatures [5][6], In pulsed laser system, the SAs plays a crucial role in the cavity photon pulsation process. When a sufficiently large number of photons circulating in the laser cavity are built, they excited the absorber and deplete the absorption, causing the transmission increase, which is known as loss saturation. The abrupt decrease of the cavity loss increases the photon life in laser cavity so the population inverted gain media can efficiently amplify the intra-cavity photons by stimulated emission. It results in the formation of burst of photons in the cavity and the output photons. However, if the absorber is easily saturated, it could stay in a saturated state during the laser built up process and cannot modulate the cavity loss. In this situation the pulsed operation can not be activated and laser delivers CW output. Notably it is due to the low saturation intensity of TI-SA so the QS-CW transition can be observed without damaging the saturable absorbers inside the laser cavity. To better understand the saturation property in the prepared TI-SA, experiments with quasi-CW pump is conducted. We turn on the pump for 10 ms and then keep it off for 6 ms repeatedly to investigate the dynamics of laser built-up and possible relaxation behaviors. Figure 3 shows the evolution of the time trace recorded by oscilloscope as the quasi-CW pump power increases. When pump power is close to the threshold value of 781 mW as shown in Fig.3(a), only slight spontaneously emitted lights are observed in the time interval when the pump is on. At quasi-CW pump power rated 2580 mW, laser pules is observed, as illustrated in Fig.3 (b). Increasing the quasi-CW power up to 4330 mW which is plotted in Fig. 3(c), the laser pulses become unstable and irregular. Given stronger pump power, as shown in Fig. 3(d-f), the laser eventually comes to a steady state of CW operation. The time required for the initial spiking to relax to the steady CW state decreases as pump power increase, which we believe should be the typical laser spiking and relaxation oscillation behavior in lasers. It infers that under very strong pump, for example, in Fig. 3(f), the saturable absorber is completely bleached and cannot provide any modulation strength to produce laser pulses and the laser output fast damped to CW state with a damping time of 150 $\mu$s. For a further justification, we compare the relaxation oscillation and laser spiking when the TI-SA is removed as shown in Fig.4 (a) which is taken under the quasi-CW pump power of 2830 mW. By taking Fourier transform of the time trace measured, the relaxation oscillation frequency and the damping time of the laser system can be obtained as 36.9kHz and 251 $\mu$s, respectively. By varying the quasi-CW pump power, the relaxation oscillation frequency increases linearly to the square root of pump power and the damping time decreases proportional the inverse of pump power and is around the level of the fluorescence lifetime as shown in Fig. 4(b) and Fig. 4(c), respectively, which well agrees with the theoretical expression derived from classical rate equations[1]. When it is compared laser dynamics at the over saturated state, the relaxation dynamics is verified.

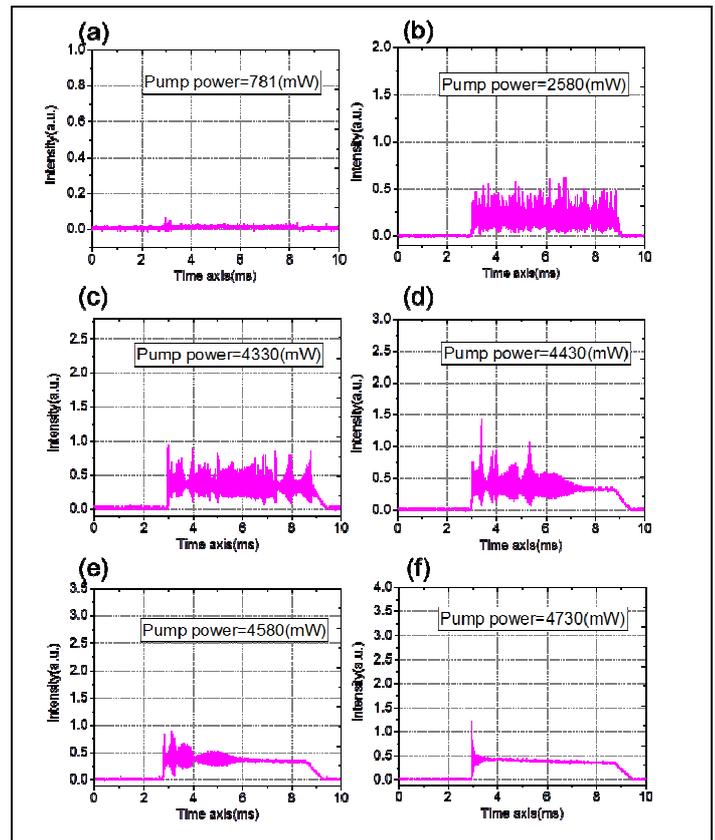

Figure 3 The evolution of the time trace recorded by oscilloscope as the quasi-CW pump power are at (a) 781 mW, (b) 2580 mW, (c) 4330 mW, (d) 4430 mW, (e) 4580 mW and (f) 4730 mW, respectively.

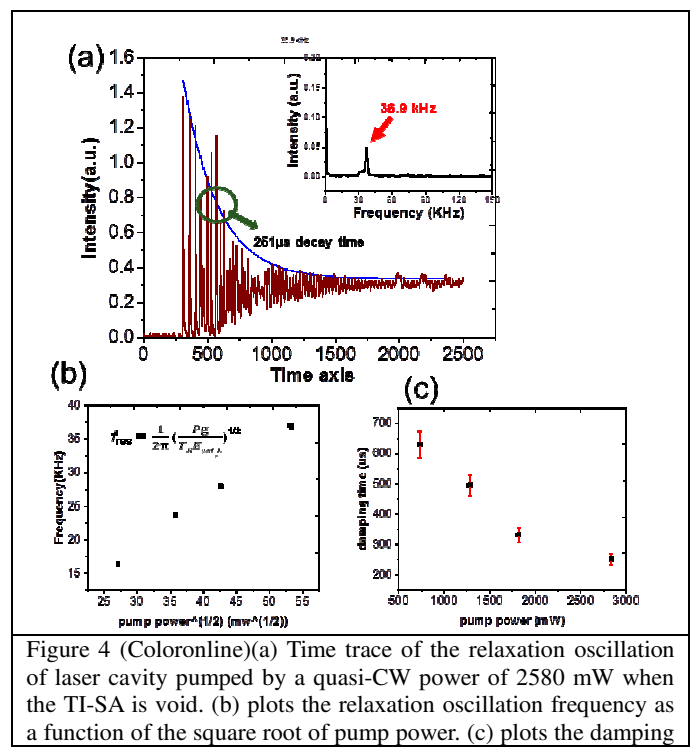

Figure 4 (Coloronline)(a) Time trace of the relaxation oscillation of laser cavity pumped by a quasi-CW power of 2580 mW when the TI-SA is void. (b) plots the relaxation oscillation frequency as a function of the square root of pump power. (c) plots the damping

## III. MODULATION INSTABILITY AND PASSIVE Q-SWITCHED LASER PULSATION IN OVER-SATURATED ABSORBER

To get further insight to the QS to CW transition in a PQS solid state laser system with low-saturation absorber, model based on the modulation instability is proposed. As a matter of fact, Powell's approach [5] is equivalent to the study of the modulation instability (MI) property in the mathematical model of coupled rate equations. MI is the central nonlinear process that delineates the energy exchange between the periodic perturbations and a homogeneous background [10]. As a result of MI, homogeneous states evolve into spatial or temporal periodic structures that emulate the character of the nonlinear process. MI therefore is an useful tool to probe the manifestation of strong nonlinear effects in nature. For the system of interest, PQS laser with a fast SA in which the absorber has a much faster recovery time than the duration of a single Q-switched pulse, it can be described as a simple intensity depending loss and the mathematical model of coupled rate equations involves only the atomic population density of gain media N and intra-cavity photon density f [3]. The MI analysis then requires small perturbations added upon the steady-state solution to obtain the linearized equations for the perturbation fields proportional to ansatz exp(qt). It can be derived that

Modulation instability (MI) is the central nonlinear process that delineates the energy exchange between the periodic perturbations and a homogeneous background[26]. As a result of MI, homogeneous states evolve into spatial or temporal periodic structures that reflects the character of the nonlinear process. It therefore serves as a manifestation of strong nonlinear effects in nature.

An example model to be considered to elucidate the MI process is nonlinear Schrödinger equation (NSE) in which the nonlinear interaction is a function of intensity $F(|u|^2)$,

$$i\frac{\partial u}{\partial Z} + \frac{1}{2}\frac{\partial^2 u}{\partial x^2} + F(|u|^2)u = 0. \quad (1)$$

This equation has the simplest solution in the form of CW plane wave wave, $u(z,x) = u_0 \exp(ipz + iqx)$ where $u_0$ is a constant and $p$, $q$ satisfy the dispersion (spatial) relation $p = -q^2/2 + F(|u_0|^2)$. This solution shows that such a plane wave of amplitude $u_0$ propagates through nonlinear medium unchanged except for acquiring an intensity-dependent phase shift.

A standard approach to identify whether this plane wave is stable against small perturbation is the linear stability analysis, which explore evolution of the perturbed solution of the form, $u = (u_0 + \epsilon(x,z))\exp(ipz + iqx)$, where $\epsilon = (u_1 + iv_1)\exp[iKz + iQx]$ represent small perturbations in complex form. To the first order of $\epsilon$, a linearized coupled equations are obtained,

$$0 = \left(-p - \frac{q^2}{2} + i\frac{d}{dz} + iq\frac{d}{dx} + \frac{1}{2}\frac{d^2}{dx^2}\right)\epsilon$$
$$+ F(|u_0|^2)\epsilon + u_0\frac{dF(|u_0|^2)}{d|u_0|^2}(u\epsilon^* + u^*\epsilon) \quad (2)$$

Solution to eq. (1) was found to yield the dispersion relation, $K_\pm = -qQ \pm Q\sqrt{Q^2/4 - F(u_0)^2}$.

In case $\frac{Q^2}{4} < F(|u_0|^2)$, $K_\pm$ are complex conjugate pair and has a negative imaginary part denoted as $g(Q) = Q\sqrt{F(u_0^2) - Q^2/4}$ that leads to the exponential growth of the periodic perturbations as illustrated in Fig. 5 and consequently changes the plane wave background to some periodic structures.

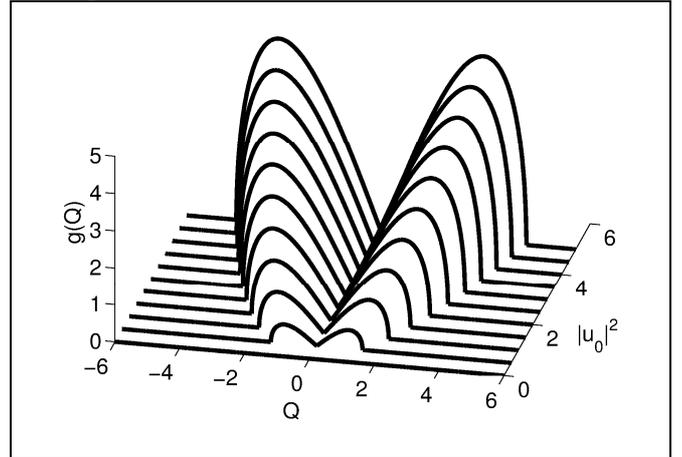

Fig. 5 Plotted is the amplitude growth rate of periodic perturbations, $g(Q) = Q\sqrt{F(u_0^2) - Q^2/4}$, as a function of background intensity $|u_0|^2$ and modulation frequency $Q$.

Dynamic cavity effect in laser systems have been studied since laser had been invented. Subtle transient behaviour has been delineated by theoretical consideration of a mathematical model of rate equations involving atomic population inversion density of gain media, intra-cavity photon density and the population dynamics of absorber if existed [1].

Passive Q-switching (PQS) is one of the mature techniques used to produce repetitive burst of coherent photons since its first demonstration by Wood and Schwarz [2] in $CO_2$ laser system. Such a pulsed operation relies on the significant saturation of the absorber inside a laser cavity. Although direct numerical simulation using the mathematical model of coupled rate equations reveals the observed repeated pulsation dynamics, Powell *et al.* proposed a perturbed approach based on linear stability analysis and reached the same conclusion in the repetition rate as the experimental works done by Karlov *et al.* [3]. Notably Powell's approach is equivalent to the study of the MI property in the mathematical model of coupled rate equations. A typical model system is the PQS laser with a saturable absorber (SA). When the absorber has a much faster recovery time [Add reference here] than the duration of a single Q-switched pulse, it can be described as a simple intensity depending loss and the mathematical model of coupled rate equations involves only the atomic population density of gain media and intra-cavity photon density write,

$$\frac{dN^*}{dt} = -c\sigma_{21}N^*\phi - \frac{N^*}{\tau_2} + P \quad (3)$$

$$\frac{d\phi}{dt} = c\sigma_{21}N^*\phi - \frac{\phi}{\tau_c} - \frac{c\sigma_a N_a^i}{1+\phi/\phi_a}\phi \quad (4)$$

in which the parameters are listed in Table 1.

| | |
|---|---|
| $N^*$ | Population inversion density of laser gain media |
| $\phi$ | Intra-cavity photon density |
| $N_a^i$ | Initial population difference of absorber |
| $\sigma_{21}$ | Gain cross section |
| $\sigma_a$ | Absorption cross section |
| $\tau_2$ | Upper level lifetime of gain media |
| $\tau_c$ | Cavity photon lifetime |
| $\phi_a$ | Saturation photon density of absorber |
| $c$ | Speed of light |
| $P$ | Pump rate |

TABLE 1. Symbols used in Eq. (3) and Eq. (4).

To a better elucidating the physical insight, we normalize the equations with a time scale $\bar{t} = t/\tau_2$ and introduce the saturation photon density, $\phi_g \equiv \frac{1}{\tau_2 \sigma_{21} c}$, unsaturated loss, $a_0 \equiv c\tau_2 \sigma_a N_a^i$ and ratio between the upper level relaxation time and cavity life time $r \equiv \tau_2/\tau_c$. The threshold population inversion can be obtained by having a positive growth rate for the intra-cavity photon density starting at zero intra-cavity photon density and it leads to an expression of the threshold population inversion density of unsaturated gain media, $N_{th}^* = (a_0 + r)\phi_g$. The pump rate can then be expressed as $P \equiv mN_{th}^*/\tau_2$ and the population inversion is feasible for $m > 1$ where the pumping over threshold denoted by $m$. The rate equations then write,

$$\frac{dN^*}{d\bar{t}} = -\frac{N^*}{\phi_g}\phi - (N^* - m(a_0 + r)\phi_g) \quad (5)$$

$$\frac{d\phi}{d\bar{t}} = \frac{N^*}{\phi_g}\phi - r\phi - \frac{a_0}{1+\phi/\phi_a}\phi \quad (6)$$

The steady state solutions of Eq. (5) and Eq. (6) are given by having $\frac{dN^*}{d\bar{t}} = \frac{d\phi}{d\bar{t}} = 0$ and a second order polynomial equation can be obtained in which the solution is the steady-state intra-cavity photon density.

$$\frac{r}{\phi_a \phi_g}\phi^2 = \left[\frac{ma_0+(m-1)r}{\phi_a} - \frac{a_0+r}{\phi_g}\right]\phi + (m-1)(a_0+r) \quad (7)$$

For $m > 1$ the only physical solution with positive photon density from the quadratic equations is,

$$\phi_{ss} = \frac{\left[\frac{ma_0+(m-1)r}{\phi_a} - \frac{a_0+r}{\phi_g}\right]}{2\frac{r}{\phi_a \phi_g}}$$

$$+ \frac{\sqrt{\left[\frac{ma_0+(m-1)r}{\phi_a} - \frac{a_0+r}{\phi_g}\right]^2 + 4\frac{r}{\phi_a \phi_g}(m-1)(a_0+r)}}{2\frac{r}{\phi_a \phi_g}}$$

(8)

and the associated steady-state atomic population inversion of gain media is

$$N_{ss}^* = \frac{m(a_0+r)\phi_g}{1+\phi_{ss}/\phi_g} \quad (9)$$

The MI analysis then requires small perturbations added upon the steady-state solutions as $N^* = N_{ss}^* + n$ and $\phi = \phi_{ss} + f$. Substituting the perturbed solution into Eq.~(3) and Eq.~(4) and neglecting high order terms of the perturbations and one obtains the linearized equations for the perturbation fields,

$$d\vec{X}/d\bar{t} = M\vec{X} \quad (10)$$

wherein $\vec{X} = [n, f]^T$ and

$$M = \begin{bmatrix} -\frac{\phi_{ss}}{\phi_g} - 1 & -\frac{N_{ss}}{\phi_g} \\ \frac{\phi_{ss}}{\phi_g} & \frac{N_{ss}}{\phi_g} - r - \frac{a_0}{1+\phi_{ss}/\phi_a} + \frac{\phi_{ss}a_0}{\phi_a(1+\phi_{ss}/\phi_a)^2} \end{bmatrix}$$

The first order system of ordinary differential equations has solutions following the form, $\vec{X} = \vec{X}_0 e^{q\bar{t}}$ and we substitute the ansatz into the Eq.(10) to get equation $q\vec{X}_0 = M\vec{X}_0$. The solution $q$ and $\vec{X}_0$ are the eigenvalue and eigenvectors of matrix $M$. The eigenvalue $q$ obeying the equation,

$$q^2 + \left(Y + \frac{\phi_{ss} - N_{ss}^*}{\phi_g} + 1\right)q +$$
$$\left[\left(\frac{\phi_{ss}}{\phi_g} + 1\right)\left(Y - \frac{N_{ss}^*}{\phi_g}\right) + \frac{\phi_{ss}N_{ss}^*}{\phi_g^2}\right] = 0 \quad (11)$$

where

$$Y = r + \frac{a_0}{1+\phi_{ss}/\phi_a} - \frac{\phi_{ss}a_0}{\phi_a(1+\phi_{ss}/\phi_a)^2}$$

To have PQS operation of repetition frequency beyond zero solution to $q$ must be a complex with negative real part and none-zero imaginary part. A negative real part ensures the growth of the perturbations as a consequence of MI. The repetition frequency of the PS pulse train is given by $f_{rep} = \Im\{q\}\tau_2^{-1}$. The repetition frequency and MI growth rates of a PQS laser system under various pumping condition are illustrated in Fig. 6 (a). It is seen that even when the pumping exceeds lasing threshold with $m > 1$, the PQS operation does not start unless $m$ goes beyond a *second threshold* value [1]. In the shaded region of Fig. 6(a) where PQS operation is possible, the repetition frequency increases from zero as a continuous-wave (CW) state with pump strength because the population inversion density burnt by the photon burst are restored at a faster speed. The time to initiate a new photon burst is shortened. On the contrary the MI growth rate decreases at stronger pumping condition because the steady-state CW background to generate MI is larger. It causes the saturated TI-SA to have weaker modulation depth to the cavity loss and the nonlinearity is therefore reduced. Notably when pumping strength goes beyond the condition $m > 1.22$, the PQS operation cease to exist because the MI growth rate is negative and the small periodic perturbations are loses its energy during the time evolution. It should results in the abrupt change from the PQS operation back to CW state in the laser system with TI-SA in contrast to the smooth transition begins at CW state to PQS operation at a relative weaker pumping. The behavior predicted by the MI analysis can be justified by performing a direct computer simulation by numerical evolutions as shown in Fig.6(b). For small pump strength, repetition frequency is low; for very strong pump the TI-SA is totally bleached and cannot provide the nonlinear modulation capability and the pulse is relaxed into a CW background.

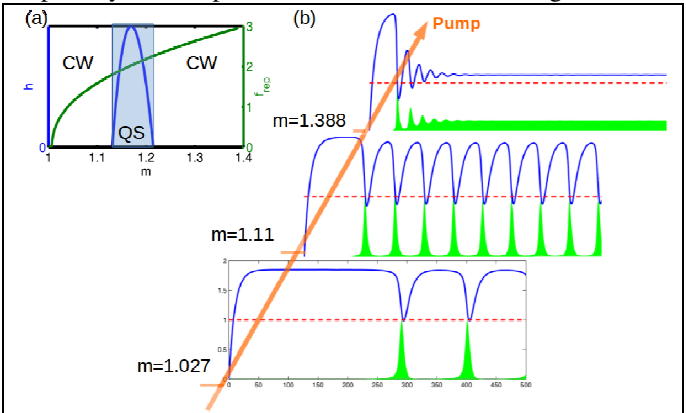

Fig. 6 (a)Repetition frequency and MI growth rate of a PQS laser system under various pumping speed above the threshold. The QS operation occurs only at the shaded regime (b)Simulated time evolution of the atomic population inversion density (in blue) and intra-cavity photon density (in green). The parameter used in the simulation are $a_0 = 8$, $r = 10$ and $\phi_g/\phi_a = 1.25$.

The expressions given in Eqs. (8,9) and Eq. (11) can be used to reveal the parameter space in which laser is operated in CW or QS mode. By further normalizing the photon density and the population inversion to the saturation photon density in the laser gain media, The laser dynamics are related only to the pumping speed over threshold, $m$, ratio of the saturation photon density of the gain media over SA, $\phi_g/\phi_a$, ratio of the upper level lifetime of laser gain media over the cavity lifetime, $r$ and the unsaturated loss coefficient. Figure 7 reveals the parameter space in terms of the ratio of saturation photon density in gain media over the SA, $\phi_g/\phi_a$ and the required pumping power to the laser gain media that supports QS operations under various conditions. It is clear that increasing $\phi_g/\phi_a$ cause the QS operation to exist only at a small region of pumping power above the second threshold. In all the three conditions plotted in Fig. 3, the increase of $\phi_g/\phi_a$ also make the second threshold to overlap the first threshold and the laser can operated at QS state once it is pumped to lase. As one compare the region in gray enclosed in dashed line and the region in black, it is seen that the larger unsaturated loss in SA produces temporal stronger modulation that goes with the temporal intensity distribution and consequently the QS operation can occurs over a larger pumping power. Moreover altering $r$ by changing the cavity lifetime as we compare the region in gray enclosed in solid line and the region in black, lower $r$ also reduces the region to have QS operation both in the possible pumping power and also in the ratio $\phi_g/\phi_a$. This can be understood by that for lower $r$, the cavity lifetime is increased and the cavity is more capable of storing optical power and therefore the SA is easier to get over-saturated.

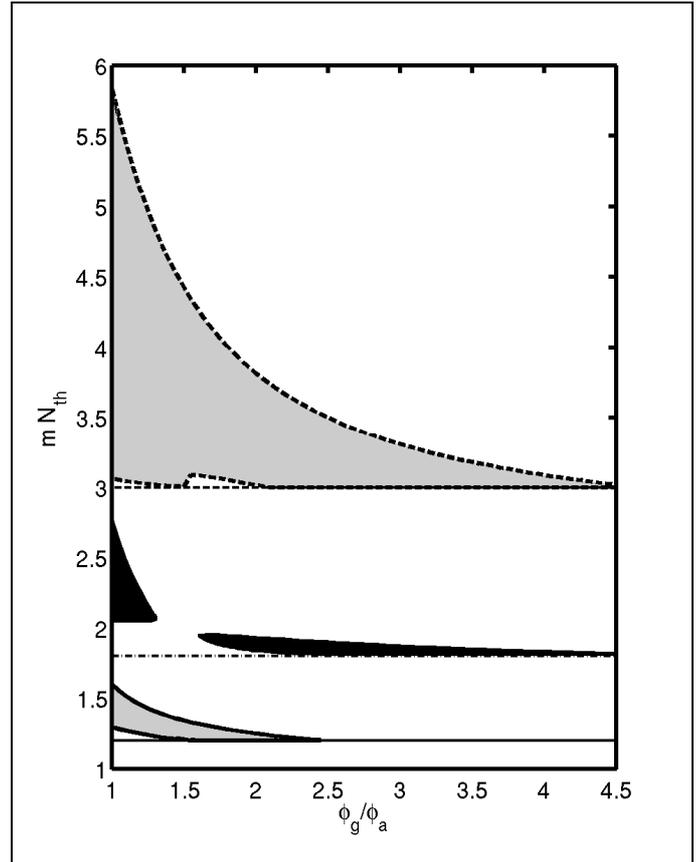

Fig. 8 Phase space in terms of the ratio of saturation photon density in gain media over the SA in the horizontal axis and the pumping speed in the vertical axis. The region in gray enclosed by dashed line, in black and in gray enclosed by solid line are for $a_0 = 20$, $r = 10$, $a_0 = 8$, $r = 10$ and $a_0 = 8$, $r = 4$, respectively. The corresponding the first pumping thresholds are plotted in dased, dashed-dotted and solid for the three conditions, respectively.

In conventional PQS DPSS, the ratio $\phi_g/\phi_a$ is about unit and the saturated intensity is typically at the order of MWs. The state of over-saturation can hardly be observed. By using the topological insulator (TI) of hexagonal $Bi_2Te_3$ single crystals with uniform morphology prepared by hydrothermal exfoliation [25] route as an intra-cavity absorber device, a saturation intensity as low as a tens to hundreds of kW can be achieved. The dynamical transition from QS to CW operation in DPSS can be observed as shown in Fig. 2(a). When the pump power is over 4 W, QS operation is terminated and CW operation is observed as illustrated in the inset of Fig. 2(a) labeled as Q-Switch state and CW state. For TI materials are known to have a much faster recovery time [25] than the duration of a single Q-switched pulse which enable us to describe contribution of the TI based saturable absorber (SA) as a simple intensity depending loss. The mathematical model of coupled rate equations involves only the atomic population density of gain media and intra-cavity photon density given by Eq.(3) and Eq. (4). Whether laser operates in CW or QS mode can therefore be predicted using Eqs. (7-11). The calculation shows that the growth rate of MI become negative when the m goes above 5 and it can qualitatively emulate the experimental observations. Moreover, by selecting the output coupling (OC) of the laser cavity to be 5%, 10% and 25%, it is seen in Fig. 8 that the QS operations sustains in the pumping power range of 223mW, 392mW and 442mW, respectively as predicted by Eq. (1) and illustrated in Fig. 7(c). Moreover, the average intra-cavity laser power at terminating pump power for the three different OC conditions are estimated to be around 3W, which reflects the complete saturation of the TI-SA.

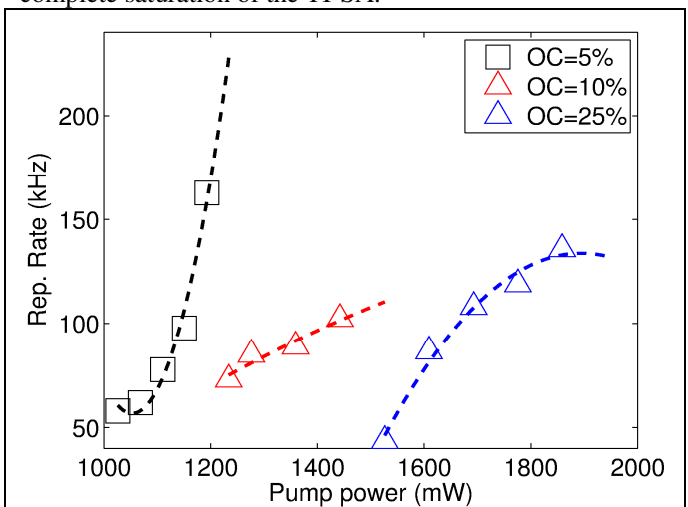

Figure. 8 (Color online) The repetition frequency of the QS pulses is plotted as a function of pumping power when the OCs of the laser cavity are 5%, 10% and 25%, respectively.

## CONCLUSION

To conclude, the reversible dynamic transition from Q-switched pulse state to continuous wave state was observed experimentally on a Topological Insulator $Bi_2Te_3$ based Passive Q-switched (QS) solid state laser and explained by modulation instability (MI) analysis based on laser rate equations. According to the MI analysis a phase space diagram in terms of pumping and the ratio of the saturation intensity of gain media over absorber reveals the parameter space in which passive QS pulse can be generated. With different photon lifetime which can be manipulated by the OC of the laser cavity, supported pomp power to form QS pulse are also experimentally justified.

ACKNOWLEDGMENT

We wish to acknowledge the support of Ministry of Science and Technology (MOST) under the contract number MOST-104-2112-M-110-001, MOST-105-2112-M-110-002.